\documentclass[superscriptaddress,aps,pra,twocolumn,showpacs,nofootinbib,longbibliography,notitlepage,floatfix]{revtex4-2}
\usepackage{amsmath,amssymb,amsthm}
\usepackage[colorlinks=true,citecolor=blue,urlcolor=blue]{hyperref}
\usepackage[pdftex]{graphicx}
\usepackage{times,txfonts}
\usepackage{braket}
\usepackage{color}
\usepackage{natbib}
\usepackage{amsmath,blkarray}
\usepackage{mathtools}
\usepackage{ulem}
\usepackage{latexsym}
\usepackage{tabularx, booktabs}
\usepackage{graphics,epstopdf}
\usepackage{graphicx}
\usepackage{float}
\usepackage{amsfonts}
\usepackage{color,soul}

\newcommand{\be}{\begin{equation}}
\newcommand{\ee}{\end{equation}}
\newcommand{\ba}{\begin{eqnarray}}
\newcommand{\ea}{\end{eqnarray}}

\usepackage{multirow}
\usepackage{appendix}
\usepackage{url}

\begin{document}
\title{Efficient broadband frequency conversion via shortcuts to adiabaticity}
\author{Koushik Paul}
\email{koushikpal09@gmail.com}
\affiliation{Department of Physical Chemistry, University of the Basque Country UPV/EHU, Apartado 644, 48080 Bilbao, Spain}
\affiliation{EHU Quantum Center, University of the Basque Country UPV/EHU, Barrio Sarriena, s/n, 48940 Leioa, Spain}

\author{Qian Kong}
\email{kongqian@shu.edu.cn}
\affiliation{International Center of Quantum Artificial Intelligence for Science and Technology (QuArtist) \\ and Department of Physics, Shanghai University, 200444 Shanghai, China
\\}

\author{Xi Chen}
\email{chenxi1979cn@gmail.com}
\affiliation{Department of Physical Chemistry, University of the Basque Country UPV/EHU, Apartado 644, 48080 Bilbao, Spain}
\affiliation{EHU Quantum Center, University of the Basque Country UPV/EHU, Barrio Sarriena, s/n, 48940 Leioa, Spain}
\begin{abstract}

The method of adiabatic frequency conversion, in analogy with a two-level atomic system, has been put forward recently and verified experimentally to achieve robust frequency mixing processes such as sum and difference frequency generation. Here we present a comparative study of efficient frequency mixing using various techniques of shortcuts to adiabaticity (STA) such as counter-diabatic driving and invariant-based inverse engineering. We show that, it is possible to perform sum frequency generation by properly designing the poling structure of a periodically poled crystal and the coupling between the input lights and the crystal. The required crystal length for frequency conversion is significantly decreases beyond the adiabatic limit. Our approach significantly improves the robustness of the process against the variation in temperature as well as the signal frequency. By introducing a single parameter control technique with constant coupling and combining with the inverse engineering, perturbation theory and optimal control, we show that the phase mismatch can be further optimized with respect to the fluctuations of input wavelength and crystal temperature  that results into a novel experimentally realizable mixing scheme.

\end{abstract}
\maketitle

\section{Introduction}
\label{sec:intro}%
 In the field of nonlinear optics, the nonlinear frequency conversion via three wave mixing process is a fundamental concept \cite{Boyd2003}, in which light of two colors is mixed in a nonlinear crystal, resulting in sum frequency or difference frequency generation (SFG or DFG) with a third color. However, the conversion efficiency of standard frequency conversion, based on quasi-phase matching (QPM) technique, is not perfect especially for broad optical signal, since the three-wave mixing processes is sensitive to the input wavelength, crystal temperature, interaction length and incidence angle. 
Remarkably, the analogy of different quantum-optical phenomena, including rapid adiabatic passage (RAP) and even stimulated Raman adiabatic passage (STIRAP) in two- or three-level atomic systems, opens new exciting possibility to control dynamics in nonlinear optical media, see recent review \cite{Suchowski}. In a specific simplification, the coupled wave equations of SFG and DFG processes in the undepleted pump approximation is analogous to time-dependent Schr\"{o}dinger equation of the interaction of light with two-level atom. Therefore, RAP with Landau-Zener scheme in frequency conversion has been suggested and also realized experimentally in aperiodically poled potassium titanyl phosphate (APPKTP) device, with high efficiency over a wide bandwidth \cite{Suchowski2009,Suchowski2011,SuchowskiPRA,Suchowski2013,karnieli2022}. In addition, the extension of STIRAP to two-process frequency conversion has been discussed in the depleted pump regime \cite{Porat2012}. Apart from adiabatic process, composite pulses are proposed to achieve efficient and boardband sum frequency \cite{Genov2014}. However, both processes require long interaction length, which shows the downside.

 In the past decade, shortcuts to adiabaticity (STA) \cite{odelin3,chenPRL} have been developed to speed up the adiabatic processes in various quantum systems \cite{STA_rev}. Among them, counter-diabatic (CD) driving \cite{demirplack1,demirplack2} (or equivalently quantum transitionless driving \cite{berry2009}) and Lewis-Riesenfeld (LR) invariant based inverse engineering \cite{LRI,mugaLRI} provides efficient ways to design the interactions that drives the system along a desired instantaneous eigenstate of the reference Hamiltonian. This approach has been extensively studied and implemented in different contemporary fields, including atomic physics \cite{expt3,masuda}, spintronics \cite{BanPRL12,YueSciRep}, quantum computation \cite{hegade2021} and  many-body state dynamics \cite{campo1,campo3}). Based on the analogy between the Schr\"{o}dinger equation and the coupled wave equation \cite{Longhi}, the STA techniques has been also exploited in optical waveguide devices, including mode conversion, directional coupler, and beam splitting \cite{Chung_2019,paul2015}. Furthermore, the optimization of STA \cite{DamesD} can be further applied for designing high coupling efficiency, robust, and short-length coupled-waveguide devices \cite{MSP}. Regarding the frequency process, the conventional CD field, one of the STA techniques, has been also first envisaged to improve the SFG process \cite{Xu16}.

In this paper, we  study extensively the shortcuts to adiabatic frequency conversion in a nonlinear aperiodically poled crystal structure by focusing on ingredients of robustness and optimality. In order to compare, we design the couplings and the phase mismatch inside the crystal using both, the CD driving and the LR invariant method to reduce the crystal length specifically for standard SFG and DFG processes and achieve efficient frequency conversion. To make physical implementation feasible, we apply the unitary transformation to find alternative coupled-mode equations that mimics the dynamics in the interaction picture. Moreover, the phase mismatch and coupling coefficients are obtained and optimized to speed up the adiabatic sum frequency generation with Landau-Zener (LZ) scheme in shorter crystal length. In order to obtain better experimental feasibility, we develop the optimization of LZ scheme in order to facilitate the frequency mixing process to be controlled by a single parameter. We show that application of STA enhances the robustness of the mixing process against the fluctuations of temperature and the input wavelength. Finally, our results are compared with the conventional frequency conversion proposed by quasi-phase matching technique to demonstrate the robustness against different crystal length and pump intensities.

The paper is organized as follows. In Sec. \ref{sec:AdSFG} we review the adiabatic SFG method followed by developing CD driving for SFG using the transitionless quantum driving approach for a single qubit in Sec. \ref{sec:CDSFG}. Sec. \ref{sec:LRI_SFG} provides the LR invariant-based engineering for SFG where, using a perturbative approach, we develop the LZ optimization based on a single control parameter. In Sec. \ref{sec:Efficiency}, we study the mixing efficiency with respect to the variation in the temperature and the signal frequency for different pump intensities and crystal length and finally we conclude in Sec. \ref{sec:conclusion}.


\begin{figure}[]
	\centering\includegraphics[width=\linewidth]{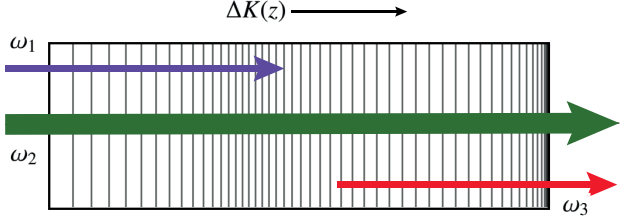}
	\caption{Schematic of controlled aperiodic structure of a poled crystal with designed continuous variation of phase mismatch $\Delta K(z)$ along the direction of propagation for realizing shortcuts to adiabatic sum frequency conversion, where  $\omega_1$,  $\omega_2$ are the signal and pump frequencies respectively whereas $\omega_1 + \omega_2 = \omega_3$ represents the idler frequency. }
	\label{fig0}
\end{figure}

\section{Adiabatic sum frequency generation}
\label{sec:AdSFG}

The nonlinear frequency mixing process for generating sum (difference) frequency corresponds to the production of an idler frequency when a QPM nonlinear crystal is subjected to a strong pump field and a relatively weak signal field, see Fig. \ref{fig0}.  
This process can be described by the coupled mode theory,
\begin{equation}
 \frac{d \tilde{A}_{1}}{dz} = -i q \tilde{A}_{3} e^{-i\Delta k z},
\qquad
 \frac{d \tilde{A}_{3}}{dz} = -i q^{\ast} \tilde{A}_{1} e^{i\Delta k z},
 \label{coupledeq2}
\end{equation}
where $\tilde{A}_{1}$ and $\tilde{A}_{3}$ are the normalized signal and idler amplitudes respectively, given by
\begin{equation}
\tilde{A}_{1} = \frac{c}{4 \omega_1} \sqrt{\frac{k_1}{\pi \chi^{(2)} {A}^{\ast}_2 }} A_1 , \quad \tilde{A}_{3} = \frac{c}{4 \omega_3} \sqrt{\frac{k_3}{\pi \chi^{(2)} {A}_2 }} A_3.
\end{equation}
The coupling coefficient, $q$ is $z$-dependent with $z$ being the propagation distance, being represented as
\begin{equation}
q(z) = \frac{4 \pi \omega_{1}\omega_{3}}{\sqrt{k_{1}k_{3}}c^{2}} \chi^{(2)} A_{2},
\end{equation}
where $A_2$ is the pump amplitude which is strong compared to signal and the idler so that the undepleted pump approximation can be assumed, $\omega_1$,  $\omega_2$ are the signal and pump frequencies respectively whereas $\omega_1 + \omega_2 = \omega_3$ represents the idler frequency. corresponding wave numbers characterizes the phase mismatch $\Delta k=k_{1}+k_{2}-k_{3}$ and $\chi^{(2)}$ represents the nonlinear susceptibility of the medium.  

Note that Eq.~\eqref{coupledeq2} is analogous to a resonant two-level quantum system coupled by complex field $q(z)=Q(z)e^{i\phi(z)}$, where $Q(z) = |q(z)|$ mimics the Rabi frequency and $\phi(z)$ being the chirping parameter \cite{Imeshev2000,Liqing10}. To study the adiabatic evolution of such a system, we introduce a unitary transformation as follows,
\begin{equation}
\quad \tilde{A}_1 = a_1 e^{-i[\Delta k  - \phi(z)]/2}, \quad \tilde{A}_3 = a_3 e^{i [\Delta k  - \phi(z)]/2}.
\label{unitary1}
\end{equation}
which turns into a rotating wave approximated Schr\"odinger like equation, 
\begin{equation}
	\label{Scheq}
i \frac{d}{dz} \begin{pmatrix} a_1 \\ a_3 \end{pmatrix} = \frac{1}{2}\begin{pmatrix} \Delta k - \dot{\phi}(z)& 2Q \\ 2Q&  \Delta k - \dot{\phi}(z)\end{pmatrix} \begin{pmatrix} a_1 \\ a_3 \end{pmatrix},
\end{equation}
with the Hamiltonian of the system being 
\begin{equation}
H(z) = \frac{\Delta K(z)}{2} \sigma_z + Q(z) \sigma_x,
\label{Ham}
\end{equation}
where, $\sigma_{x,y,z}$ represents the well-known Pauli matrices. To achieve the adiabatic SFG, the most important parameter is $\Delta K(z)$. The sweeping process of quasi phase matching requires designing the crystal with appropriately structured poling period such that the sweeping occurs from $-\Delta_0$ to $\Delta_0$. Here we follow the traditional LZ model to choose the the phase mismatch with constant coupling $Q = Q_0$ and $\Delta K(z)  = \Delta_0 - \alpha z$. the spatial evolution of such a system is governed by the so called adiabatic condition which can be calculated by using the dressed state picture. 
\begin{figure}
\centering\includegraphics[width=\linewidth]{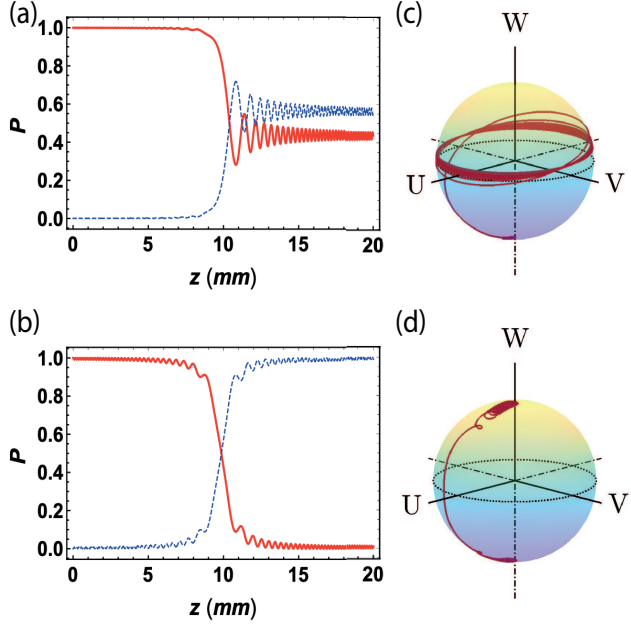}
\caption{Conversion of modes along the direction of propagation of the crystal, where (a) incomplete conversion with adiabatic condition is violated with $I_P = 60~ \text{MW}/\text{cm}^2$ and (b) conversion is complete as the adiabatic condition is satisfied with $I_P = 360 ~ \text{MW}/\text{cm}^2$, (c) and (d) respective Bloch vector trajectories.
}
\label{fig1}
\end{figure}
\begin{equation}
\label{condition}
C_{ad} = \left|\frac{Q_0 \partial_z \Delta K}{(Q_0^{2}+{\Delta K^{2}})^{3/2}}\right| \ll 1,
\end{equation}
Clearly, the adiabaticity of the evolution is dictated by the choice of the phase mismatch $\Delta K(z)$, which can be taken as \cite{Liqing10}
\begin{equation}
\label{mismatch}
    \Delta K(z) = \Delta K_0 + \delta K + \delta K_{\Lambda}(z).
\end{equation}
$\Delta K_0 = k_1 + k_2 - k_3$ is the primary phase mismatch, $\delta K$ is the contribution due to group velocity mismatch. The last term characterizes QPM which can be expressed as $2 \pi/ \Lambda(z)$ . Here $\Lambda(z)$ represents the poling period of the crystal. Also the wave vector $k_i = 2 \pi n_i/\lambda_i$ where, $n_1$, $n_2$ and $n_3$ are the refractive indices corresponding to the respective frequencies. Note that, all the refractive indices are taken along the extraordinary axis of the polarization and these are sensitive the temperature variations which can be quantified in terms of the Sellmeier equation \cite{Fan87}.

The LZ slope $\partial_z \Delta K = -\alpha$ also depends on the choice of the crystal length and the range of the poling period in Eq.~\eqref{mismatch}. As the  spatial evolution occurs in a crystal of a constant length, so one can design only a fixed $\alpha$ for a particular sample. In fact, during the SFG, the adiabaticity maintains using the variation of poling period along the crystal length. For instance, if the initial poling period is $\Lambda_i$ and $\Lambda_f$ be the final with $L$ being the total length of the crystal, then $\alpha = (\Delta K_i - \Delta K_f)/L $ where $\Delta K_{i,f} = 2 \pi / \Lambda_{i,f}$. To study such a system, we consider APPKTP with periodicity $ \Lambda_i$ varies from $16.2~ \mu\text{m}$  to $ 14.6~ \mu\text{m}$ along the direction of propagation. Also, the offset $\Delta_0$ is determined from the first two terms of Eq.~(\ref{mismatch}) and can be adjusted accordingly to drive the $\Delta K(z)$ from $-\Delta_0$ to $\Delta_0$. Therefore, the adiabaticity can only be controlled externally by the choice of the $Q_0$ which explicitly depends upon the pump intensity, $I_P = c \epsilon_0 A_2^2/2$. 

Figure~\ref{fig1} depicts the transfer of power from $\omega_1$ mode to $\omega_3$. We choose our crystal size to be 20~mm for the adiabatic SFG with $\chi^{(2)} = 32~\text{pm/V}$. The signal, pump and idler wavelengths are chosen as $\lambda_1 = 1535~ \text{nm}$, $\lambda_2 = 1064 ~\text{nm}$, $\lambda_3 = 643~\text{nm}$ \cite{Suchowski2009,SuchowskiPRA}. For pump intensity $I_P = 60 ~\text{MW}/\text{cm}^2$, the adiabatic condition fails, see Eq.~\eqref{condition}, with $C_{ad} > 1$ and complete mode conversion can not be achieved as shown in Fig.~\ref{fig1} (a) as well as in the Bloch vector trajectory in Fig.~\ref{fig1} (b). Whereas for $I_P = 360 ~ \text{MW}/\text{cm}^2$, $C_{ad} < 1$, satisfies Eq.~\eqref{condition} successfully which results into perfect mode transfer, demonstrated in Fig.~\ref{fig1} (c) and (d). It is evident that, like all adiabatic processes, the adiabatic SFG is also slow as it requires relatively large crystal length as well as large pump intensity $ > 360 ~\text{MW}/\text{cm}^2$ to achieve greater efficiency.

\section{Counter-diabatic sum frequency generation}
\label{sec:CDSFG}
 In this section, we focus on the CD driving for sum frequency generation, which has been developed in Ref. \cite{Xu16}. The instantaneous eigenmodes of $H(z)$ (\ref{Ham}) can be written as 
$(\ket{n_+(z)}, \ket{n_-(z)})^T = U(\vartheta(z))^\dagger(\ket{0}, \ket{1})^T$,
where $U(\vartheta(z))$ represents a unitary rotation with the mixing angle being $\vartheta(z)=2 \tan^{-1}(2 Q_0/\Delta K)$. And $(\ket{0}, \ket{1})^T$ represents the basis modes characterizing $a_1$ and $a_3$ respectively and the interaction Hamiltonian can be expressed in terms of $\ket{n_{\pm}(z)}$ basis by using the unitary transformation,
\begin{equation}
H_a(z) = U^{\dagger}(\vartheta(z)) H_(z)U(\vartheta(z)) - iU^{\dagger}(\vartheta(z))\dot{U}(\vartheta(z)).
\end{equation}
According to the CD driving \cite{demirplack1,demirplack2}  (or the transitionless quantum driving \cite{berry2009}) it is always possible to construct a driving Hamiltonian, which cancels out the non-adiabatic part $iU^{\dagger}(\vartheta(z))\dot{U}(\vartheta(z))$. Addition of a driving term in $H_a(z)$ drives the system exactly along the adiabatic path even beyond the adiabatic in Eq.~\eqref{condition}. The driving Hamiltonian, $H_1(z)$ is constructed from the instantaneous eigenstates which is Hermitian and purely off-diagonal in nature, can be written in adiabatic basis as \cite{berry2009},
\begin{equation}
{H}_1 = i \sum_{\pm}|\partial_{z} n_{\pm} \rangle \langle n_{\pm}|,
\label{adham}
\end{equation}
from which for our system, the Hamiltonian ${H}_1$ finally takes the following form:
\begin{equation}
{H}_{1}= \frac{1}{2}\begin{pmatrix}
0& i \dot{\vartheta}&\\
-i \dot{\vartheta}& 0
\end{pmatrix},
\label{admat}
\end{equation}
In principle, the total Hamiltonian ${H_{eff}}= {H(z)} + {H}_1$ can transfer $a_1$ to $a_3$ in a fast adiabatic-like way, which means the state evolves from $a_1$ to $a_3$ along the instantaneous eigenstate of Hamiltonian ${H}_0$ within short propagating distance, not satisfying the adiabatic condition (\ref{condition}). Taking into account the physical implementation, we further simplify the total Hamiltonian ${H_{eff}}= {H}(z) + {H}_1$ by using the concept of multiple Schr\"{o}dinger picture \cite{MSP}, and finally obtain
\begin{equation}
{H}_{eff}=\begin{pmatrix}
 \frac{\Delta K(z) - \dot{\varphi}}{2}&  Q_{eff}&\\
Q_{eff}& - \frac{\Delta K(z) - \dot{\varphi}}{2}
\end{pmatrix},
\label{effham}
\end{equation}
where $ \varphi (z)=\tan^{-1} (\dot{\vartheta}/Q_0)$ and $Q_{eff}=(\dot {\vartheta}^2/4 +Q_0^2)^{1/2}$.
\begin{figure}
\centering
\includegraphics[width=\linewidth]{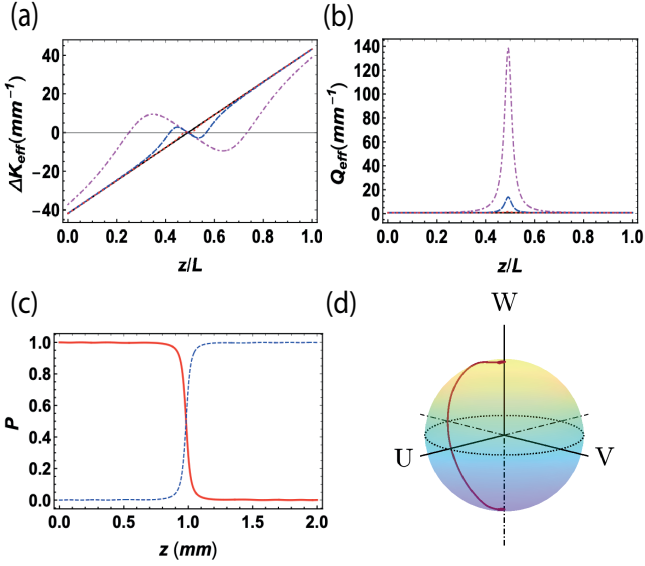}
\caption{ (a) Profile of required $\Delta K_{eff}$ for the application of CD driving for different crystal lengths, i.e., $L = 20~\text{mm}$ (dotted red), $L = 2 ~\text{mm}$ (dashed blue), $L = 0.2~\text{mm}$ (dot-dashed magenta), and adiabatic one $L = 200 ~\text{mm}$  (solid black). (b) Nature of additional coupling required for complete mode transfer in different lengths. (c) Conversion of modes along the direction of propagation of the crystal using the CD driving for crystal length $2 mm$ with $I_P = 60~ \text{MW}/\text{cm}^2$,  and (d) respective Bloch vector trajectory.}
\label{fig2}
\end{figure}
This method promises that there is a possibility for achieving frequency conversion in very small crystal length. However that requires modification in both the phase mismatch and the coupling. This is evident from Fig.~\ref{fig2} (a) and (b), which shows the required modification of $\Delta K(z)$ and $Q_0$. It shows that the smaller the crystal size, the more drastic the modification is required to achieve complete mode transfer. But if the modifications are achieved, complete mode transfer is guaranteed in infinitesimally small crystal length. As show in Fig.~\ref{fig2} (c), the mode transfer is complete even for $L = 2 ~\text{mm}$ and the Bloch vector in Fig.~\ref{fig2} (d) shows the required path is smaller compared to the adiabatic one which refers to the smaller crystal length.
The results convincingly show that the CD approach for SFG is much superior when in terms of the crystal size and the coupling strength. However, the more we decrease the crystal length, the changes in the effective phase mismatch are more rapid. as the system consists of a single APPKTP crystal only, it could be challenging to design inside a very small crystal length with existing poling methods. Moreover, the amplitude of the effective coupling increases rapidly with the decreasing crystal length. This can be achieved using an applied external field along the energy transfer region of the crystal \cite{Xu16}. Perhaps an alternative method would be to 
focus the pump to the center of the crystal, by designing or further optimizing the focused pump beam characteristics \cite{zhou2014generation}.


\section{Optimal sum frequency generation}
\label{sec:LRI_SFG}
\begin{figure}
	\centering
	\includegraphics[width=\linewidth]{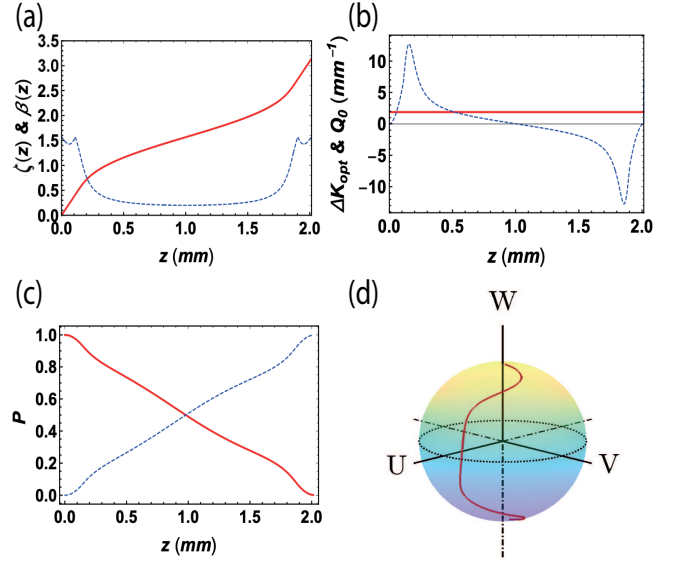}
	\caption{\label{para_opt} (a) $z$ dependence of $\zeta$ (solid red) and $\beta$ (dashed blue), obtained from Eqs.~(\ref{gam}) and (\ref{beta}), with the boundary conditions as satisfied by the eigenstates of LR invariant i.e., $\zeta(0) = 0$ and $\zeta(L) = \pi$.  (b) The coupling constant $Q_0$ (solid red) and optimal $\Delta K_{opt}$ (dashed blue) designed from Eq. (\ref{optDel}) by using the LZ optimization of the LR invariant engineering. (c) The corresponding conversion of modes along the direction of propagation of the crystal and (d) respective Bloch vector trajectory. Parameters: $c_1=-1.47$ for crystal length $2~\text{mm}$ with $I_P = 360~\text{MW}/\text{cm}^2$.}
	\label{fig3}
\end{figure}
 In this section, we will opt for the LR invariant-based engineering for STA \cite{LRI,mugaLRI} in SFG.  Although the CD approach shows robust and fast SFG, from the implementation viewpoint it poses significant difficulties in practical implementation. In general, inverse engineering method is based on designing the coupling and the phase mismatch simultaneously from the imposed boundary conditions for dynamical modes of LR invariant. However, we aim to focus on an optimization of the phase mismatch with respect to a constant coupling, making it easier implementation by removing the requirement of additional coupling. To this end, one needs first to construct the LR invariant, which is generally chosen in a parameterized form, yielding \cite{Ruschhaupt,DamesD}
\begin{equation}
I(z) =\frac{I_0}{2}(\sin \zeta(z) \cos \beta(z) {\sigma}_x - \sin \zeta(z) \sin \beta(z) {\sigma}_y + \cos \zeta(z) {\sigma}_z),~~~~~
\label{inv}
\end{equation}
where $I_0$ is an arbitrary parameter and has the dimension of coupling coefficient. The invariant equation which is to be satisfied is given by:
\begin{equation}
\frac{dI(z)}{dz} = i\frac{\partial I(z)}{\partial z} - [H(z),I(z)].
\label{inv_eq}
\end{equation}
Here $H(z)$ is the original Hamiltonian. From above equations we obtain the following conditions for invariance:
\begin{subequations}
\begin{align}
\dot{\zeta}(z) &= 2 Q(z) \sin \beta(z),\label{gam} \\
\dot{\beta}(z)  &= -\Delta K(z) + 2Q(z) \cot \zeta (z) \cos \beta(z).\; \; 
\end{align}
\label{invariance}
\end{subequations}
The LR invariant posses a different set of eigenmodes compared to the Hamiltonian which do not coincide in general. These eigenmodes can be written in parametric form as:
\begin{equation}
\ket{\Phi^+} = \begin{pmatrix} 
\cos({\frac{\theta(z)}{2}})e^{-i\beta} \\ \sin({\frac{\theta(z)}{2}})
\end{pmatrix}, ~
\ket{\Phi^-} = \begin{pmatrix} 
\sin({\frac{\theta(z)}{2}}) \\ -\cos({\frac{\theta(z)}{2}})e^{i\beta}
\end{pmatrix} .
\end{equation}
In principle, the solution of model equations (\ref{Scheq}), resembling Sch\"{o}dinger equation, can be written as the superposition of eigenmodes of dynamical invariant, see below.  And the instantaneous eigenmodes $\ket{\Psi_{\pm}(z)}$ are related to $\ket{\Phi_{\pm}(z)}$ by LR phase, given by $
\ket{\Psi^{\pm}} = \ket{\Phi^{\pm}} e^{i \gamma_{\pm}(z)}$,
with the LR phase being deduced as
\begin{equation}
\dot{\gamma}_{\pm} = \pm \bigg( \dot{\beta} + \frac{\dot{\theta} \cot \beta}{\sin{\theta}} \bigg).
\end{equation}

\subsection{Landau-Zener optimization}
\label{sec:LZ_opt}

To optimize the invariant-based shortcut according to the LZ scheme, we find an optimized profile for $\Delta K_{opt}$ for a constant $Q(z)$ in order to make it more feasible for practical situations. Considering $Q(z) = Q_0$, and from Eqs. (\ref{invariance}) we get \cite{DingML}
\begin{equation}
\label{optDel}
\Delta K_{opt} = -\frac{\ddot{\zeta}(z)}{2 Q_0 \bigg( 1-\frac{\dot{\zeta}(z)^2}{4 Q_0^2} \bigg)^2} + 2 Q_0 \cot \zeta(z) \bigg( 1-\frac{\dot{\zeta}(z)^2}{4 Q_0^2} \bigg)^2.
\end{equation}
One should note that, from the above equation, with this choice an additional constraint comes to the system, thus it is obvious that $\Delta K_{opt}$ only depends on $\zeta$. For the optimization, we follow the perturbative approach for the systematic error \cite{Ruschhaupt,DamesD}. The error with respect to the signal wavelength $\lambda_1$ in the phase mismatch term from Eq.~\eqref{mismatch} is \cite{tseng4} 
 %
%
\begin{equation}
\delta_{\lambda_1} \approx -\frac{2 \pi n_1 \delta \lambda_1}{\lambda_1^2},
\end{equation}
which presents the perturbative error, described by $ H' =\delta_{\lambda_1}\sigma_z/2$. The probability for the system to be found in a particular mode $\ket{\Phi^+}$ can be written as \cite{DamesD},
 \begin{equation}
 P_+(z) = 1 - \left(\frac {\delta_{\lambda_1}} {2}\right)^2\bigg| \int_0^L \bra{\Phi^-} \sigma_z \ket{\Phi^+} dz \bigg|^2 ,
 \end{equation}
 from which can define the error sensitivity as follows \cite{Ruschhaupt}:
 \begin{equation}
 q_\Delta = - \frac{1}{2} \bigg| \frac{\partial^2 P_+}{\partial {\delta_{\lambda_1}}^2}  \bigg|^2.
 \end{equation}
Now  $ P_+$ has to be unity in order to maintain the system in $ \ket{\Phi_+}$, which follows that 
\begin{equation}
\int_0^L dz \sin \zeta(z) \exp(i m(z)) \rightarrow 0,
\label{integration}
\end{equation}
where $m(z) = 2 \gamma_+-\beta$.  To make the above integral we expand $m(z) \approx m(\zeta)$ in terms of Fourier series \cite{DamesD},
\begin{equation}
m(\zeta) = 2 \zeta +c_1 \sin 2\zeta + ... + c_n \sin 2 n \zeta + ...
\label{coeff}
\end{equation}
It is straightforward to calculate that 
\begin{equation}
\beta = \cot ^{-1} \bigg( \frac{1}{2 M \sin\zeta}  \bigg),
\label{beta}
\end{equation} 
with $M = \frac{1}{2}\frac{dm}{d\zeta}$. Combining all Eqs. (\ref{invariance}) and (\ref{beta}), we can obtain 
\begin{equation}
\label{bound}
 Q_0 L  = \int_{0}^{\pi} \sqrt{1+ 4M^2 \sin^2 \zeta } d\zeta \geq \pi,
\end{equation}
which sets the bound for crystal length with the maximum value of $Q_0$ allowed.
It should be noted that with this optimization, unlike conventional invariant-based approach, we loose the freedom to design the parameters $\zeta$ and $\beta$ which characterizes the invariant itself. One can find $\zeta$ by solving Eq.~(\ref{invariance}) using Eq.~(\ref{beta}) and  design an optimal $\Delta K_{opt}$ from Eq.~\eqref{optDel}. The only parameter one can chose are the Fourier coefficients in order to nullify the integral Eq.~(\ref{integration}). Accordingly we can also obtain the intensity of the pump field in order to maintain a constant $Q_0$, as follows,
\begin{equation}
I_2(\lambda_1) =   \bigg( \frac{Q_0^2 c \epsilon_0}{32 \chi ^{(2)}} \bigg)\lambda_1 \lambda_3 n_1 n_3.
\end{equation}
 Similar  optimization can be made with respect to the other parameters as well. For instance, the temperature dependence of the wavelength can be studied by the optimization with respect to the refractive index along the extraordinary axis.  
 In Fig.~\ref{fig3} (a), we have plotted $\beta$ and $\zeta$ which are obtained by solving the Eq.~\eqref{gam} and Eq.~\eqref{beta}. Like the conventional LR invariant method, where one has the freedom to design the coupling and the phase mismatch using the boundary conditions, the $\zeta$ shows similar behavior (varies from $0$ to $\pi$), but $\beta$ changes drastically in order to maintain the constant coupling. Here we have used only one Fourier coefficient in Eq.~\eqref{coeff}, i.e., $c_1$ in order to find optimal $\Delta K_{opt} $ as shown in Fig.~\ref{fig3} (b). Moreover, as it turns out, the optimal length depends on the two parameters only which are basically $c_1$ and $Q_0$. Since in the LZ optimization, $Q_0$ is a constant with $I_2 = 360~\text{MW}/\text{cm}$, for a fixed crystal length of $2~\text{mm}$ the only degrees of freedom we have is the choice of $c_1$. In Fig.~\ref{fig3} (c), we choose $c_1=-1.47$ which results in a complete mode conversion along a fixed path, shown in the corresponding Bloch vector trajectory in Fig.~\ref{fig3} (d). The product $Q_0L$ is always constant, see Eq. (\ref{bound}), for a particular value of $c_1$ and it approaches to $\pi$ when the higher order terms in Eq.~\eqref{coeff} are considered.
 %
%
%
%
%
\begin{figure}
\centering\includegraphics[width=\linewidth]{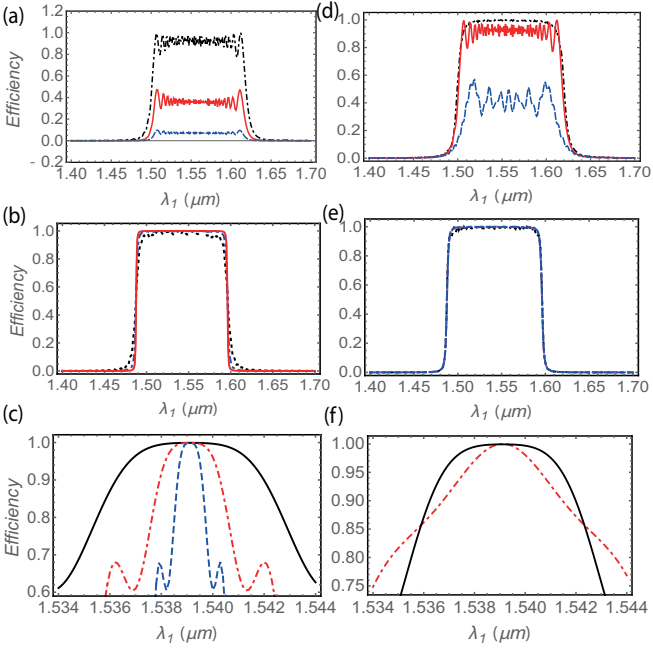}
\caption{ Conversion efficiency of modes with respect to the variation of signal wavelength for different pump intensities with $I_P = 10~\text{MW}/\text{cm}^2$ (blue dashed), $I_P = 60~\text{MW}/\text{cm}^2$ (solid red) and $I_P = 360~\text{MW}/\text{cm}^2$ (black dashed-dotted ) in (a), (b) and (c); and for different crystal length with  pump intensity $I_P = 360~\text{MW}/\text{cm}^2$, $L=2~\text{mm}$ (blue dashed), $L = 10 ~\text{mm}$ (solid red), $L = 20~\text{mm}$ (black dashed-dotted) in (d), (e) and (f) respectively. Here for comparison, (a,d) present adiabatic SFG, (b,e) presents CD driving, and (c,f)  presents the optimal SFG designed by inverse engineering. }
\label{fig4}
\end{figure}

\section{Efficiency}
\label{sec:Efficiency}

\begin{figure}
\centering\includegraphics[width=\linewidth]{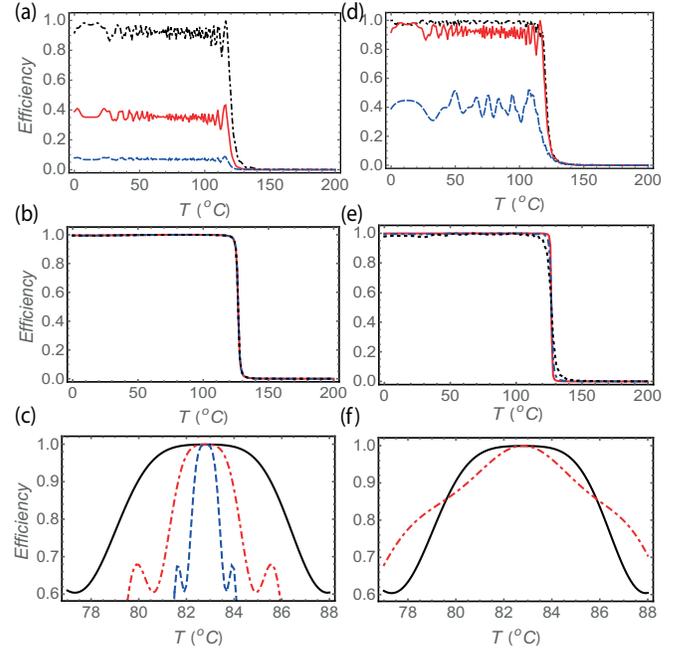}
\caption{Conversion efficiency of modes with respect to the variation of temperature for different pump intensities with $I_P = 10~\text{MW}/\text{cm}^2$ (blue dashed), $I_P = 60~\text{MW}/\text{cm}^2$ (solid red) and $I_P = 360~\text{MW}/\text{cm}^2$ (black dashed-dotted) in (a), (b) and (c); and for different crystal length with with pump intensity $I_P = 360~\text{MW}/\text{cm}^2$, $L=2 \text{mm}$ (blue dashed), $L = 10 ~\text{mm}$ (solid red), $L = 20 \text{mm}$ (black dashed-dotted) in (d), (e) and (f) respectively. Here for comparison, (a,d) present adiabatic SFG, (b,e) presents CD driving, and (c,f)  presents the optimal SFG designed by inverse engineering. } 
\label{fig5}
\end{figure}

The robustness of the aforementioned STA inspired SFG methods can be demonstrated by studying the conversion efficiency against the variation of externally controllable parameters, where the efficiency is defined as $|a_3(L)|^2/|a_1(0)|^2$. In Fig.~\ref{fig4} and Fig.~\ref{fig5} , we present a comparative study of the conversion efficiency with respect to the input signal wavelength and the crystal temperature for different peak pump amplitude and crystal length. 
\subsection{Dependence on the wavelength}
\label{sec:wavelength}

 Figure~\ref{fig4} shows the efficiency of the SFG with respect to the variation of signal wavelength. for adiabatic case, it shows broadband nature and thereby robust against the wavelength variation. In Fig. \ref{fig4} (a) variation of  efficiency for different pump intensity for the adiabatic case is clarified. An efficiency value close to unity can be achieved when the pump intensity is more than $360~\text{MW}/\text{cm}^2$. It is also critically dependent on the crystal length for the same poling period variation. As shown in Fig.~\ref{fig4} (d), the efficiency decreases with the decreasing crystal length with almost zero for $L=2~\text{mm}$ even when pump intensity is around $360~\text{MW}/\text{cm}^2$.

  In Fig.~\ref{fig4} (b) and (e), we examine the efficiency when the SFG is assisted by the CD driving which also  expetedly exhibit the broadband and even smoother efficiency curve. Unlike the adiabatic case, the efficiency profile is insensitive to the variation of the pump amplitude. This is mainly due to the fact that the additional coupling compensates for the requirement of the extra coupling strength for mode conversion. Moreover, the required strength of the additional coupling is higher for smaller crystal length which makes, as Fig.~\ref{fig4} (e) shown, the efficiency profile is constant with respect to different crystal length as well. 

However in case of the LZ optimal SFG, one can not compare the variation against the variation of signal wavelength and crystal length separately as $Q_0 L$ constitutes a constant quantity. Fig.~\ref{fig4} (c) depicts variation for different pump intensities where the profile is broader as the pump intensity becomes stronger and smaller crystal length (solid black). Also in Fig.~\ref{fig4} (f), efficiency is shown for two different $c_1$ values. For $c_1 = -1.47$, where we see the efficiency profile is broader compared to $c_1 = -0.2$ for $I_P = 360~\text{MW}/\text{cm}^2$. 

\subsection{Dependence on the temperature}
\label{sec:temperature}

The efficiency of the SFG also depends on the crystal temperature as the refractive index of birefringent crystals are highly dependent on temperature. Since the phase mismatch, see Eq.~(\ref{mismatch}), is a function of refactive index as well, the entire frequency conversion process becomes temperature dependent. 
Generally the refractive indices for a particular wavelength in the APPKTP crystal is determined by the Sellmeier equation, given by \cite{Fan87}
\begin{equation}
n^2 = A + \frac{B}{1-C \lambda^2} - D \lambda^2,
\end{equation} 
Constants A, B, C and D are taken from Ref. \cite{Fan87}. The temperature dependence is obtained conventionally as (for KTP crystal) \cite{Emanueli_sell}
\begin{equation}
\Delta n (\lambda,T)= n_1(\lambda) (T -25^o C) +n_2(\lambda) (T -25^o C) ^2,
\end{equation}
with
\begin{equation}
n_{1,2}(\lambda) = \sum_{m=0}^{3} a^{1,2}_m/\lambda^m.
\end{equation}
Here $a^1_m$ and $a^2_m$ are constants (see Ref. \cite{Emanueli_sell}). The temperature dependence  of the efficiency in Fig.~\ref{fig5} also shows broadband feature, depending on the pump intensity and the crystal length. For $L = 20 \text{mm}$ efficiency is higher for the higher pump intensity, see also Fig.~\ref{fig5}, but decreases when the crystal length as well as the pump intensity is decreased. For CD driving, however, these variations are eliminated and a smoother profile regardless of the crystal length and the pump intensity. Also for the LZ optimization, the variation in efficiency improved for higher pump amplitude and when $c_1 = -1.47$. 
\section{Conclusion}
\label{sec:conclusion}
In conclusion, we have studied the SFG process in an APPKTP crystal using the STA methods. We have reviewed the adiabatic SFG scheme which is extremely robust with respect to the parameter variations. However it requires large crystal length and relatively strong pump pulse to  achieve complete mode conversion. On the contrary, the STA based approaches such as the CD driving and LR invariant approach can obtain robust SFG in much shorter crystal dimensions. Application of CD driving requires modifications in the poling structure of the crystal which can be easily obtained using modern fabrication techniques for chirped quasi phase matched crystals \cite{Charbonneau08,tehranchi2008,Suchowski,Descloux15}.  In principle using CD driving one can achieve SFG in crystals with infinitesimally small length. although in reality this may be limited due to the requirement of extremely large additional coupling. For instance, to achieve the mode conversion in $0.2~ \text{mm}$, the additional coupling strength is around $140~\text{mm}^{-1}$, for which the required intensity would be very high compared to APPKTP crystals damage threshold of $500~ \text{MW}/\text{cm}^2$. However, this could be remedied by using external field along the energy conversion region of the crystal \cite{Xu16}. Although robust, the CD driving may pose significant difficulties regarding the implementation as it requires spatial modification of both the coupling and the phase mismatch simultaneously. As a potential solution, we further propose LZ optimization, by combining  LR invariant and perturbation theory, for designing the sample crystal. Following the perturbative approach, an optimal phase mismatch can be obtained by choosing the coupling as constant. This can significantly reduce the required crystal length as well as can eliminate the requirement of the additional coupling. However, it is evident that LZ optimization is not as robust compared to the adiabatic and CD approach. This is mostly due to the introduction of additional constraint. With a constant value of coupling, we have a precise value of a crystal length which guarantees mode conversion in a relatively narrow range of variation in signal wavelength and temperature. Further studies and experimentation in this direction may provide more insights to the SFG process and improve understanding of the nonlinear frequency mixing process. For instance, the variation in temperature changes the refractive index of the crystal according to the Sellmeier equation which in turn changes the phase mismatch condition itself. Therefore, it is also possible to achieve control over the phase mismatch for the STA based frequency mixing using a suitable temperature profile \cite{Rozenberg19}. Moreover, one can combine the optimally robust STA in nonlinear quantum system to the case beyond depleted pump regime \cite{Suchowski,karnieli2022}. Regarding the physical implementation, one can optimize the focused pump beam with respect to beam characteristics, i.e., focusing parameter and spatial pattern, which is also worthwhile to pursue in the future work.
%
\section*{Acknowledgements}
This work is partially supported by EU FET Open Grant  EPIQUS (899368),  QUANTEK project (KK-2021/00070),  the Basque Government through Grant No. IT1470-22, the project grant 
PID2021-126273NB-I00 funded by MCIN/AEI/10.13039/501100011033 and by
“ERDF A way of making Europe” and ``ERDF Invest in
your Future”  and NSFC (12075145) . X.C. acknowledges the Ram\'on y Cajal program (RYC-2017-22482).
%
\bibliography{main}




\end{document}